# Nanocomposite NiO:Au hydrogen sensors with high sensitivity and low operating temperature

M. Kandyla[a*], C. Chatzimanolis-Moustakas[a,b], E.P. Koumoulos[b], C. Charitidis[b], and M. Kompitsas[a]

[a]*National Hellenic Research Foundation, Theoretical and Physical Chemistry Institute, 48 Vasileos Constantinou Avenue, 11635 Athens, Greece*
[b]*National Technical University of Athens, School of Chemical Engineering, 9 Heroon Polytechniou Street, 15780 Zografou, Greece*

**Abstract**

We present results on the development of nanocomposite NiO:Au thin-film hydrogen sensors, which are able to detect hydrogen concentrations as low as 2 ppm in air, operating at low temperatures in the range 125 – 150$^o$C. Thin NiO films were sputter-deposited on oxidized silicon substrates. The structural, morphological, and nanomechanical properties of the films were investigated with respect to post-deposition annealing. Au nanoparticles were added on the NiO surface via pulsed laser deposition and the films were tested as hydrogen sensors before and after Au deposition. The performance of the NiO films as hydrogen sensors improved significantly in the presence of Au nanoparticles on the surface. The detection limit (lowest detectable hydrogen concentration) decreased by two orders of magnitude, while the response time also decreased by a factor of three.

* Corresponding author:

M. Kandyla  
National Hellenic Research Foundation  
Theoretical and Physical Chemistry Institute  
48 Vasileos Constantinou Avenue  
11635 Athens, Greece  
Tel.: +30 210 7273826, Fax: +30 210 7273794  
E-mail: kandyla@eie.gr

# 1. Introduction

Hydrogen is widely used in the chemical, petroleum, and metallurgical industries, as well as in power station cooling. Furthermore, the use of hydrogen finds increasing interest lately, due to its projection as the clean fuel of the future and as an important potential energy source. In addition to hydrogen fuel cells, which are already employed in specialized applications, hydrogen is expected to be mass produced and distributed in the near future for passenger vehicles and aircrafts, as well as a city gas. Dangers associated with hydrogen include high permeability through many materials, flammability (lowest explosion limit is 40,000 ppm in air), and lack of odor, taste, and color, which renders it undetectable by human senses [1]. Therefore, smart hydrogen sensors with high sensitivity and low power consumption are essential in order to achieve safe and efficient processing of hydrogen on a massive scale.

Gas sensor technologies vary depending on the mechanism of operation, such as resistive, electrochemical, catalytic, optical, and mechanical, among others. Each technology presents important advantages and certain disadvantages, depending on the application. Resistive gas sensors detect changes in the electrical resistance of a material in the presence of an analyte gas [1],[2]. The advantages of resistive gas sensors include low cost, high sensitivity, and wide operating temperature range. Metal-oxide thin films have been successfully employed as resistive sensors due to their electrical response in the presence of a reducing or oxidizing gas. Metal oxides such as $SnO_2$ [3], ZnO [4], and $TiO_2$ [5], among others, have shown very good hydrogen sensing properties, such as high sensitivity, fast response, and long-term stability, combined with low-cost and flexible production, as well as simplicity in their use. One of the drawbacks of metal-oxide sensors, is that they have to be heated during operation in order to promote the reaction with the analyte gas. In general, the operating temperature of such sensors ranges between 180 and 450$^{o}$C [1], increasing the power consumption of the devices.

The advent of nanotechnology provides new structures and materials for gas sensing applications. Metal-oxide nanorods, nanowires, nanobelts, and nanodiscs [6],[7],[8],[9], core-shell nanostructures [10], heterojunction nanofibers [11], and carbon nanotubes [12] are some examples of novel hydrogen nanosenors. New types of materials such as organic-inorganic hybrids [13] and graphene [14] have also been used recently for hydrogen sensing. The advantages of nanostructures include large active surface area, high charge carrier mobility, high sensitivity, and low fabrication costs.

Residual stresses may be developed during the micro/nano fabrication process of sensor devices, which can seriously affect their operating performance and reliability, as they can cause rupture and/or delamination of thin films and nanostructures [15]. The mechanical and tribological behavior of sensing devices is of critical importance in determining long-term stability and reliability. Good maintenance of the mechanical and tribological properties of sensors can significantly impact their commercialization. Nanoindentation is a useful tool for the control of stresses for successful and reliable sensor operation. The variation of the nanomechanical/nanotribological response of sensors, as determined by nanoindentation, should be kept within a narrow range of values.

In this work, we present results on the fabrication of resistive nanocomposite NiO:Au hydrogen sensors, which are able to detect hydrogen concentrations as low as 2 ppm in air, operating at temperatures in the range 125 – 150$^{o}$C. Additionally, the structural, morphological, and nanomechanical properties of the sensors are investigated with respect to post-fabrication annealing. Even though NiO is a widely employed metal oxide, with excellent chemical stability [16], ease of fabrication, and applications ranging from electrochromism [17] to fuel cells [18], its use as a gas sensor material is limited. This is because NiO is a p-type semiconductor due to nickel vacancies [19], contrary to most other metal oxides, which are n-type semiconductors due to oxygen vacancies. It has been shown that n-type materials present higher response to analyte gases than p-type materials because surface effects dominate the resistive behavior of the former [20]. The most promising results reported so far for resistive NiO hydrogen sensors include the detection of concentrations down to 500 ppm, operating at the $300 - 650^{o}C$ temperature range, by thin NiO films [21],[22]. Also, NiO films with thin Pt overlayers or embedded Au nanoparticles have been tested as resistive hydrogen sensors, detecting concentrations of 300-500 ppm while operating at lower temperatures, in the $150 - 420^{o}C$ range, due to the promoting role of the metallic elements [23],[24]. Pure NiO films have been shown to operate at $125^{o}C$ for higher hydrogen concentrations of 3000 ppm [25]. Films of hollow NiO nano-hemispheres were employed as hydrogen sensors, detecting 200 ppm of hydrogen at $300 - 400^{o}C$ [26]. By employing NiO films with Au nanoparticles on the surface, we are able to detect two orders of magnitude lower hydrogen concentrations at lower operating temperatures, compared with concentrations and temperatures found in the literature, therefore improving the detection limit and reducing the power consumption of the devices. Thus, NiO performance is now comparable to the performance of n-type metal-oxide sensors, making NiO an attractive material for hydrogen sensing. Even though previous works in this field mainly employ embedded metallic nanoparticles in the bulk of the sensor film, the results presented in this paper are obtained with a very small amount of Au nanoparticles, which exist only on the sensor surface.

## 2. Materials and methods

Thin NiO films were deposited by DC reactive magnetron sputtering from a nickel target (76 mm diameter, 99.95% purity) on oxidized silicon substrates kept at room temperature. A mixture of oxygen and argon flows was used, controlled by mass flow controllers. The total gas pressure was kept at 0.6 Pa and the oxygen partial pressure was kept at 0.18 Pa. A sputtering power of 600 W was used. To investigate the effect of annealing, some samples were post-annealed at 400 - 600$^{o}$C for 1 h in nitrogen (annealed samples), while others were used for measurements without annealing (as-deposited samples).

Au nanoparticles were deposited on the NiO surface by pulsed laser deposition. The deposition took place in a vacuum chamber evacuated to a pressure of 10$^{-5}$ mbar. Au targets were irradiated for 60 s by a Q-switched Nd:YAG laser system (9 ns pulse duration, 10 Hz repetition rate), operating at 355 nm. The NiO - Au target distance was 50 mm and the NiO samples were kept at 100$^{o}$C during Au deposition.

The surface morphology of the samples was investigated with the aid of a Philips Quanta Inspect Scanning Electron Microscope (SEM), equipped with an Energy Dispersive Spectrometer (EDS). A Theta D5000 X-ray diffractometer (XRD) with Cu K$_\alpha$ radiation was employed for structural characterization.

Nanoindentation measurements were performed on the NiO samples with a Hysitron Tribolab Nanomechanical Test Instrument, which allows the application of loads from 1 to 30,000 μN on the surface of the sample and records the displacement of the surface as a function of the applied load with high load resolution (1 nN) and high displacement resolution. The obtained curve is called load-displacement curve.

The nanomechanical test instrument is equipped with a Scanning Probe Microscope, in which a sharp probe tip moves in a raster scan pattern across the sample surface using a three-axis piezo positioner. For each sample, a total of 10 indents with a spacing of 50 μm in a clean area environment with 45% humidity and 23°C ambient temperature were averaged to determine the sample's mean hardness, $H$, and Young's modulus, $E$. In order for the test instrument to operate under closed loop load or displacement control, a feedback control option was used. All nanoindentation measurements were performed with the standard three-sided pyramidal Berkovich probe, with an average radius of curvature of about 100 nm [27], 40 s loading and unloading time, and 3 s of holding time, to avoid residual viscoelasticity [28],[29]. Prior to indentation, the area function of the indenter tip was measured in fused silica, a standard material for this purpose [30].

Based on the half-space elastic deformation theory, $H$ and $E$ values for the NiO samples can be extracted from the experimental data (load-displacement curves) using the Oliver-Pharr method [31]. The derived expressions for calculating the elastic modulus, $E$, from indentation experiments are based on Sneddon's [32],[33] elastic contact theory:

$$E = \frac{S\sqrt{\pi}}{2\beta\sqrt{A}} \qquad (1),$$

where $S$ is the unloading stiffness (equal to the slope of the load-displacement curve at the beginning of unloading), $A$ is the projected contact area between the tip and the substrate, and $β$ is a constant that depends on the geometry of the indenter ($β$=1.167 for a Berkovich tip [31]). The nanoindentation hardness, $H$, is given by:

$$H = \frac{P_m}{A} \qquad (2),$$

where $P_m$ is the peak (maximum) applied load and the projected contact area, $A$, is calculated as:

$$A(h_c) = 24.5h_c^2 + \alpha_1 h_c + \alpha_{1/2} h_c^{1/2} + \ldots + \alpha_{1/16} h_c^{1/16} \qquad (3),$$

where

$\alpha_1 = -1.2396 \times 10^4$, $\alpha_{1/2} = 8.0499 \times 10^5$, $\alpha_{1/4} = -7.2931 \times 10^6$, $\alpha_{1/8} = 1.7166 \times 10^7$,

$\alpha_{1/16} = -1.068 \times 10^7$ \qquad (4)

and

$$h_c = h_m - \varepsilon \frac{P_m}{S_m} \qquad (5).$$

Here, $h_m$ is the total penetration displacement of the indenter at peak load, $P_m$, and $\varepsilon$ is an indenter geometry constant, equal to 0.75 for a Berkovich indenter [30],[31].

The NiO samples were tested as resistive hydrogen sensors in a home-built sensor setup, before and after Au deposition. During measurements, the samples were heated inside an aluminum chamber and their temperature was continuously recorded by a thermocouple. A flow of hydrogen, controlled by a flowmeter, was allowed in the chamber, which was filled with air at atmospheric pressure. The hydrogen concentration in air was calculated from the hydrogen partial pressure in the chamber, measured by an MKS Baratron gauge. In order to achieve low hydrogen concentrations, to the single ppm level, hydrogen was mixed with dry nitrogen in a premixing chamber, thus achieving dilution factors below $10^{-2}$. With a constant bias voltage of 1 V, the current through the samples was recorded in real time by a Keithley 485 picoammeter. The sensor response, $S$, was calculated by:

$$S = (R_g - R_o)/R_o \qquad (6),$$

where $R_g$ is the electric resistance of the sample in the presence of hydrogen and $R_0$ is the resistance of the sample in air.

## 3. Results and discussion

*3.1 Structural properties*

Figure 1a shows X-ray diffractograms of an as-deposited NiO film and of NiO films annealed at 400, 500, and 600°C after sputtering. The curves have been vertically shifted for clarity. We observe that the as-deposited film is amorphous, as it does not present any diffraction peaks, while annealing induces partial crystallization in the material. With increasing annealing temperature, the films become more and more crystalline, as indicated by the intensity of the XRD peaks. The amorphous phase is reduced as the annealing temperature increases, since more energy is supplied for crystallite growth, thus resulting in an improvement of the crystallinity of the films. There are three common peaks in the diffractograms of the annealed films, at $2\theta = 37.4°$, $43.3°$, and $63°$. Even though the cubic phase of NiO leads to diffraction planes at the same positions as the rhombohedral phase of NiO, making it difficult to distinguish between the two phases from XRD data, NiO films deposited with similar conditions to those employed in this work were found to be rhombohedral [34]. Therefore, we attribute the XRD peaks to crystallization of the films at the (101), (012), and (110) orientations of the rhombohedral phase, respectively. The average size of the NiO grains is estimated from the Scherrer equation at 15 - 20 nm. A weak peak at $56.1°$, which appears only for the film annealed at 500°C, is most probably created by the (311) plane of the silicon substrate [35], since there is no known XRD peak at $56.1°$ for either the cubic or the rhombohedral phase of NiO.

Figure 1b shows X-ray diffractograms of the NiO samples after the deposition of Au nanoparticles, where the curves have been again vertically shifted for clarity. The film annealed at 600°C was not employed as a hydrogen sensor, therefore we did not deposit Au nanoparticles on this sample. The NiO peaks corresponding to the

(101) and (012) orientations are still present and a new peak appears at $2\theta = 38.4^o$, which represents the (111) plane of the cubic phase of Au. From the Scherrer equation we estimate the average size of Au nanoparticles to be 65 - 77 nm. Even though in Fig. 1a the (012) NiO orientation peak is more intense than the (101) orientation peak, the opposite is true after Au deposition for both films annealed at 400°C and 500°C, as we can see in Fig. 1b. Given the very small thickness of the NiO films, which is approximately 10 nm as we show below in section 3.2, we attribute this effect to compressive stress induced by the deposition of Au, which creates a small change in the structural properties of the films.

*3.2 Surface morphology and composition*

Figure 2a shows an SEM image of a NiO film, before Au deposition, in side view. From this image we measure the thickness of the film to be 10 nm. All NiO films employed in this work have the same thickness. Figures 2b - 2d show SEM images of the as-deposited NiO film and films annealed at 400 and 500°C, after Au deposition. From the XRD data, presented in section 3.1, we estimate the average grain size of NiO to be 15 – 20 nm and the average size of Au particles at 65 - 77 nm. Even though it is not possible to resolve these features with the SEM magnification employed in Figs. 2b – 2d, these images provide a view of the surface morphology of the sensor films. The μm-size clusters that appear in Figs. 2b – 2d are probably particulates from the target material (Ni), which form on the surface of the NiO films during deposition, as it is often the case with sputtering.

Figure 3 shows the EDS spectrum of the NiO film annealed at 400°C, after Au deposition. We observe the presence of Ni and Au peaks, as well as the presence of a Si peak, originating from the substrate on which the NiO samples were deposited. Because the NiO films are only 10 nm thick, the magnitude of the Ni and Au peaks is much lower than the magnitude of the Si peak. Similar EDS spectra were recorded for the other NiO films after Au deposition.

*3.3 Nanomechanical properties*

Load-displacement curves, obtained by nanoindentation, are shown in Fig. 4 for the as-deposited NiO film, as well as for NiO films annealed at 400 and 500$^o$C. From these curves and Eqs. 1–5, the hardness, $H$, and elastic modulus, $E$, of the films are calculated and plotted in Fig. 5. These hardness and modulus values are not absolute values, since the films are so thin (~10 nm) that substrate effects can become important. However, based on the fact that all parameters are kept identical for all three NiO films (*e.g.*, indenter tip roundness, applied force protocol, substrate material, thickness) a direct comparison of the nanomechanical integrity of these thin film-substrate systems is possible. We observe the hardness of the NiO films decreases after annealing at 400$^o$C, while annealing at 500$^o$C induces the opposite effect. Also, the NiO film annealed at 400$^o$C shows increased elastic modulus, compared with the as-deposited and annealed at 500$^o$C samples. We note that even though the XRD data in Fig. 1 indicate gradual crystallization of the films upon annealing, they are not sufficient to predict the nanomechanical properties of the films. Indeed, from Fig. 1a we deduce a monotonic relation between annealing temperature and crystallinity, which does not follow for the hardness and elastic modulus. Therefore, nanoindentation provides complementary information for determining the optimum annealing temperature for thin-film sensors.

The resistance of the films to wear and their elasticity are determined by the ratios $H/E$ and $H^3/E^2$, respectively, which are presented in Fig. 6. The ratio of hardness/elastic modulus, $H/E$, is of significant interest in tribology. High ratio of hardness to elastic modulus is indicative of good wear resistance in a disparate range of materials, such as ceramics, metals, and polymers [36]. As we can see in Fig. 6, the NiO film annealed at 500°C shows enhanced resistance to wear, as indicated by *H*/*E*, as well as high elastic behavior under contact, as indicated by $H^3/E^2$. On the contrary, the NiO film annealed at 400°C shows reduced resistance to wear and plastic behavior, compared with the as-deposited film. These results, combined with the results on hardness and elastic modulus, indicate that the nanomechanical properties of sputtered NiO films can be controlled by careful selection of the post-deposition annealing temperature.

*3.4 Hydrogen sensing*

Figure 7 shows response curves for hydrogen sensing in air for the NiO sample annealed at 400°C, before and after deposition of Au nanoparticles. The response *S* is calculated according to Eq. 6. Because NiO is a p-type semiconductor, its electric resistance increases in the presence of hydrogen, which is a reducing gas [23],[37]. This increase is due to the reaction of hydrogen with oxygen, which is adsorbed on the surface of NiO, and the subsequent release of $H_2O$ vapor and free electrons, which recombine with the holes inside the material. Because the holes are the majority charge carriers in NiO, the reduction in the number of holes results in a resistance increase. Figure 7a shows the response of the NiO sample to hydrogen before Au nanoparticle deposition on the surface, while Fig. 7b shows the response of the sample after Au deposition. Both curves were taken at 125°C operating temperature, which is one of the lowest operating temperatures for NiO sensors reported in the literature. We observe that the deposition of Au nanoparticles allows for the detection of two orders of magnitude lower hydrogen concentrations compared to the pure NiO sample. Additionally, the response time of the sensor (defined as the time interval between 10% and 90% of the total signal change) decreased from ~15 min to ~5 min after Au deposition. Overall, the performance of the sensor is significantly improved by the presence of Au nanoparticles on the surface.

The NiO films were tested as hydrogen sensors for three different operating temperatures (125$^o$C, 130$^o$C, and 150$^o$C) and various hydrogen concentrations, before and after Au deposition. The results are summarized in Fig. 8, which shows the maximum response, *S*, for each temperature and hydrogen concentration combination for the NiO sample annealed at 400$^o$C. The response of the sensor increases with increasing operating temperature, because the adsorption-desorption kinetics, which affect the sensor performance, depend on the operating temperature [38][39]. For the same reason, higher operating temperature allows for the detection of lower hydrogen concentrations, as shown in Fig. 8a. On the other hand, it is known that if the temperature is too high, the oxidation reaction of the analyte gas (here hydrogen) proceeds so rapidly before the gas reaches the sensor surface that the gas concentration seen by the sensor decreases significantly and the sensitivity of the sensor decreases as well [37],[40]. Additionally, for high temperatures the desorption rate of hydrogen from the sensor surface becomes significant and competes with hydrogen sensing [25]. Clearly, in this work we are below the critical temperature for hydrogen and the response of the sensor improves with the temperature increase. Finding the optimum operating temperature for the detection of hydrogen is beyond the scope of this paper, which concentrates mainly on the development of highly efficient, low-power sensors. Comparing Fig. 8a with Fig. 8b we note that the deposition of Au nanoparticles on NiO improved the sensor performance for all temperatures. Table 1 lists the detection limit (lowest detectable hydrogen concentration) for each temperature with and without Au on the NiO surface. The presence of Au nanoparticles allowed for the detection of 1-2 orders of magnitude lower hydrogen concentrations, pushing the sensor detection limit to the few-ppm level.

Au nanoparticles act as catalysts, lowering the energy barrier for dissociation of hydrogen and adsorbed oxygen molecules to atomic species in their vicinity [37],[41]. Therefore, highly activated atomic hydrogen and oxygen is produced and interact with each other more efficiently, resulting in a detectable change in the electric resistance of NiO even for very low hydrogen concentrations. In the presence of Au nanoparticles, oxygen is not only adsorbed on the NiO surface but also on the surface of the nanoparticles. The interaction of these oxygen species with hydrogen produces additional free electrons in the nanoparticles, which exchange charge carriers with the NiO grains, in order to maintain neutrality. Therefore, additional free electrons are injected from the Au nanoparticles to NiO and contribute to the increase of the electric resistance of the material in the presence of hydrogen. As a result, the Au nanoparticles act not only as catalysts but they also increase the effective sensor area significantly. The combination of the above mechanisms leads to the observed orders of magnitude improvement of the sensor performance in the presence of Au nanoparticles.

The effect of post-deposition annealing on sensor performance was studied by comparing the response of the as-deposited NiO film with the film annealed at 400$^o$C. Au nanoparticles were deposited on both NiO films with identical conditions and the samples were subsequently tested as hydrogen sensors. The results are shown in Fig. 9 for two different operating temperatures. We observe that both samples were able to detect equally low hydrogen concentrations (down to a few ppm), however for both temperatures the as-deposited sample shows increased response compared to the annealed sample. The gas sensing performance of metal oxides depends on various parameters, including grain size, defects, and oxygen-adsorption properties. Annealing can induce an increase in grain size due to agglomeration, which reduces the porosity and active surface area of the film, deteriorating the sensing performance. Additionally, annealing affects the defect density of the sensing material. As mentioned in the introduction, NiO is a conductor due to the presence of lattice vacancies. During the sensing process, hydrogen interacts not only with the adsorbed oxygen on the surface of NiO, but also with the defect states of the film [39]. Annealing of the NiO samples improves their quality and decreases the defect density, therefore their performance as sensors declines. Similar effects have been observed in ZnO sensors, where sensitivity was found to be proportional to defect density [38],[42] and annealing had a detrimental effect on sensitivity [39].

## 4. Conclusions

We have presented data on the performance of nanocomposite NiO:Au thin films as hydrogen sensors. NiO films were sputter-deposited on oxidized silicon substrates and Au nanoparticles were subsequently added on the NiO surface via pulsed laser deposition. The as-deposited NiO samples were amorphous and started crystallizing with post-deposition annealing. The Au nanoparticles were found to be crystalline directly after deposition. Depending on the annealing temperature, the hardness, resistance to wear, and elasticity of the films increased or decreased compared to the as-deposited samples. This indicates that the nanomechanical properties of the films, which are essential for their long-term stability and reliability, which in turn affect their potential for commercialization, can be controlled by post-deposition annealing. The performance of the NiO films as hydrogen sensors improved significantly in the presence of Au nanoparticles on the surface. The detection limit decreased by two orders of magnitude, while the response time also decreased by a factor of three. Hydrogen concentrations in air as low as 2 ppm were detected after Au deposition. These results were obtained for operating temperatures at the lower end of what is reported in the literature. In order to improve gas selectivity, it is also possible to operate these sensors at higher temperatures, optimized for hydrogen detection. Au nanoparticles act as catalysts and also increase the effective sensor area, resulting in significant improvement of the NiO sensing performance. Even though post-deposition annealing was found to improve the crystallinity of the NiO films, it proved detrimental for their sensing properties because it decreases the active surface area and/or the density of defects, which act as hydrogen absorption sites. NiO:Au nanocomposites are promising hydrogen sensor devices, satisfying the need for high sensitivity and low power consumption, which will allow them to ensure the safe and efficient processing of hydrogen on a massive scale.

**Acknowledgements**

We would like to thank Professor I. Hotovy for the preparation of NiO samples by DC sputtering and XRD measurements.

**References**

[1] T. Hubert, L. Boon-Brett, G. Black, U. Banach, Sensor. Actuat. B 157 (2011) 329-352.

[2] M. Kandyla, C. Pandis, S. Chatzandroulis, P. Pissis, I. Zergioti, Appl. Phys. A 110 (2013) 623-628.

[3] Y.-H.Choi, S.-H. Hong, Sensor. Actuat. B 125 (2007) 504-509.

[4] N. Al-Hardan, M.J. Abdullah, A.A. Aziz, Appl. Surf. Sci. 255 (2009) 7794-7797.

[5] A.Z. Sadek, J.G. Partridge, D.G. McCulloch, Y.X. Li, X.F. Yu, W. Wlodarski, *et al.*, Thin Solid Films 518 (2009) 1294-1298.

[6] O. Lupan, G. Chai, L. Chow, Microelectr. J. 38 (2007) 1211-1216.

[7] L.L. Fields, J.P. Zheng, Y. Cheng, P. Xiong, Appl. Phys. Lett. 88 (2006) 263102.

[8] J.M. Baik, M.H. Kim, C. Larson, C.T. Yavuz, G.D. Stucky, A.M. Wodtke, *et al.*, Nano Lett. 9 (2009) 3980-3984.

[9] A. Wei, L. Pan, W. Huang, Mater. Sci. Eng. B 176 (2011) 1409-1421.

[10] N. Singh, A. Ponzoni, R.K. Gupta, P.S. Lee, E. Comini, Sensor. Actuat. B 160 (2011) 1346-1351.

[11] Z. Wang, Z. Li, J. Sun, H. Zhang, W. Wang, W. Zheng, *et al.*, J. Phys. Chem. C 114 (2010) 6100-6105.

[12] W. Wongwiriyapan, Y. Okabayashi, S. Minami, K. Itabashi, T. Ueda, R. Shimazaki, *et al.*, Nanotechnology 22 (2011) 055501.

[13] L. Renard, H. Elhamzaoui, B. Jousseaume, T. Toupance, G. Laurent, F. Ribot, *et al.*, Chem. Commun. 47 (2011) 1464-1466.

[14] W. Wu, Z. Liu, L.A. Jauregui, Q. Yu, R. Pillai, H. Cao, *et al.*, Sensor. Actuat. B 150 (2010) 296-300.

[15] V.P. Tsikourkitoudi, E.P. Koumoulos, N. Papadopoulos, E. Hristoforou, C.A. Charitidis. J. Optoelectron. Adv. M. 14 (2012) 169-175.

[16] H. Kumagai, M. Matsumoto, K. Toyoda, M. Obara, J. Mater. Sci. Lett. 15 (1996) 1081-1083.

[17] Z. Jiao, M. Wu, Z. Qin, H. Xu, Nanotechnology 14 (2003) 458.

[18] X. Chen, N.J. Wu, L. Smith, A. Ignatiev, Appl. Phys. Lett. 84 (2004) 2700.

[19] D. Adler and J. Feinleib, Phys. Rev. B 2 (1970) 3112-3134.

[20] N. Barsan, C. Simion, T. Heine, S. Pokhrel, U. Weimar, J. Electroceram. 25 (2010) 11-19.

[21] H. Gu, Z. Wang, Y. Hu, Sensors 12 (2012) 5517-5550.

[22] H. Steinebach, S. Kannan, L. Rieth, F. Solzbacher, Sensor. Actuat. B 151 (2010) 162-168.

[23] I. Hotovy, J. Huran, P. Siciliano, S. Capone, L. Spiess, V. Rehacek, Sensor. Actuat. B 103 (2004) 300-311.

[24] E.D. Gaspera, M. Guglielmi, A. Martucci, L. Giancaterini, C. Cantalini, Sensor. Actuat. B 164 (2012) 54-63.

[25] A.M. Soleimanpour, Y. Hou, A.H. Jayatissa, Sensor. Actuat. B 182 (2013) 125-133.

[26] N.G. Cho, I.-S. Hwang, H.-G. Kim, J.-H. Lee, I.-D. Kim, Sensor. Actuat. B 155 (2011) 366-371.

[27] C.A. Charitidis, Int. J. Refract. Met. H. 28 (2010) 51-70.

[28] G. Feng and A.H.W. Ngan, J. Mater. Res. 17 (2002) 660-668.

[29] H. Bei, E.P. George, J.L. Hay, G.M. Pharr, Phys. Rev. Lett. 95 (2005) 045501.

[30] M. Troyon and L. Huang, Surf. Coat. Tech. 201 (2006) 1613-1619.

[31] W.C. Oliver and G.M. Pharr, J. Mater. Res. 7 (1992) 1564-1583.

[32] I.N. Sneddon, Math. Proc. Cambridge 44 (1948) 492-507.

[33] R.B. King and T.C. O'Sullivan, Int. J. Solids Struct. 23 (1987) 581-597.

[34] I. Hotovy, J. Huran, L. Spiess, J. Mater. Sci. 39 (2004) 2609-2612.

[35] G.B. Tong and S.A. Rahman, Solid State Science and Technology 12 (2004) 47-52.

[36] A. Leyland and A. Matthews, Surf. Coat. Tech. 177 (2004) 317-324.

[37] S.R. Morisson, Sensor. Actuat. 12 (1987) 425-440.

[38] M.-W. Ahn, K.-S. Park, J.-H. Heo, J.G. Park, D.W. Kim, K.J. Choi, *et al.*, Appl. Phys. Lett. 93 (2008) 263103.

[39] O. Lupan, V.V. Ursaki, G. Chai, L. Chow, G.A. Emelchenko, I.M. Tiginyanu, *et al.*, Sensor. Actuat. B 144 (2010) 56-66.

[40] N. Yamazoe, Y. Kurokawa, T. Seiyama, Sensor. Actuat. 4 (1983) 283-289.

[41] T. Korotcenkov, L.B. Gulina, B.K. Cho, S.H. Han, V.P. Tolstoy, Mater. Chem. Phys. 128 (2011) 433-441.

[42] L. Liao, H.B. Lu, J.C. Li, C. Liu, D.J. Fu, Y.L. Liu, Appl. Phys. Lett. 91 (2007) 173110.

**Figure captions**

Figure 1. X-ray diffractograms of (a) as-deposited and annealed (at 400, 500, and 600$^o$C) NiO films and (b) NiO films with Au nanoparticles.

Figure 2. SEM images of (a) a NiO film before Au deposition in side view, (b) as-deposited NiO, (c) NiO annealed at 400$^o$C, and (d) NiO annealed at 500$^o$C. (b) – (d) images were obtained after Au deposition.

Figure 3. EDS spectrum of the NiO film annealed at 400$^o$C, after Au deposition.

Figure 4. Load-displacement curves for as-deposited NiO film and NiO films annealed at 400 and 500$^o$C.

Figure 5. (a) Hardness and (b) elastic modulus of the as-deposited NiO film and NiO films annealed at 400 and 500$^o$C, as a function of the displacement.

Figure 6. Correlation of (a) $H/E$ and (b) $H^3/E^2$ with displacement.

Figure 7. Hydrogen sensing in air under dynamic flow conditions obtained with the NiO film annealed at 400$^o$C (a) before and (b) after Au deposition.

Figure 8. Summary of sensing results for different operating temperatures and hydrogen concentrations in air, obtained (a) before and (b) after the deposition of Au nanoparticles on NiO annealed at 400$^o$C.

Figure 9. Sensor response of as-deposited and annealed at 400°C NiO films after Au deposition, at (a) 130°C and (b) 142°C operating temperatures.

**Table caption**

Table 1: Hydrogen detection limit achieved with NiO before and after Au deposition, for various operating temperatures.

**Table 1**

|                    | Detection limit (ppm) |         |
| ------------------ | --------------------- | ------- |
| **Temperature (°C)** | **Without Au**        | **With Au** |
| 125                | 123                   | 2.7     |
| 130                | 117                   | 2.6     |
| 150                | 50                    | 2       |

**Figures**

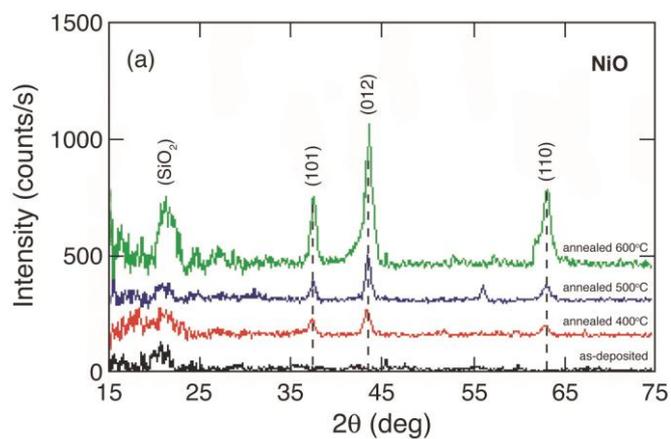

Figure 1a

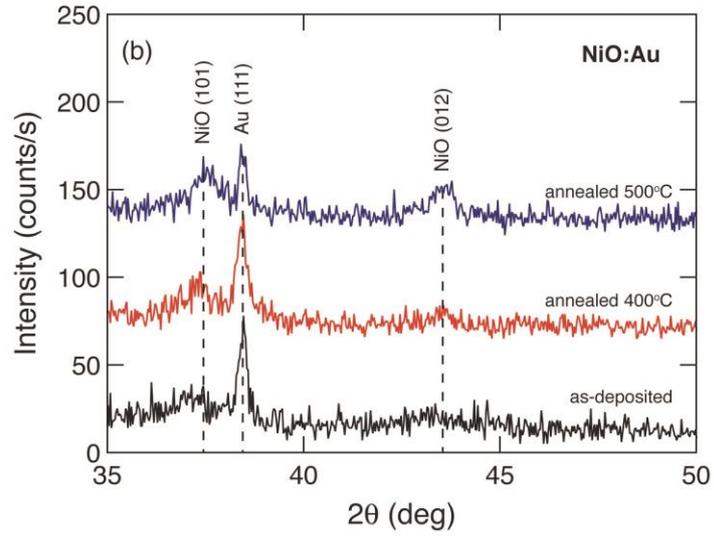

Figure 1b

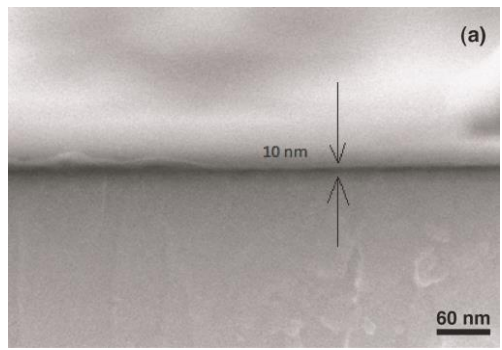

Figure 2a

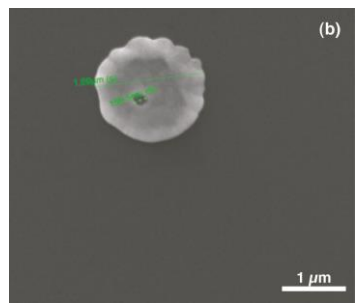

Figure 2b

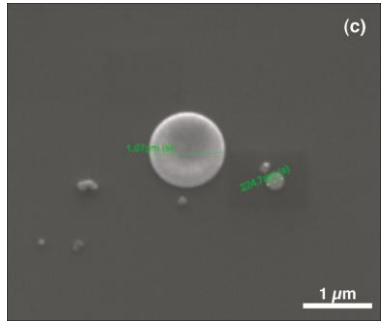

Figure 2c

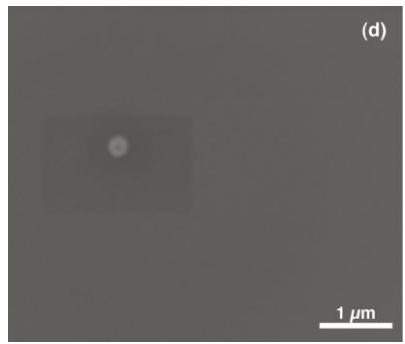

Figure 2d

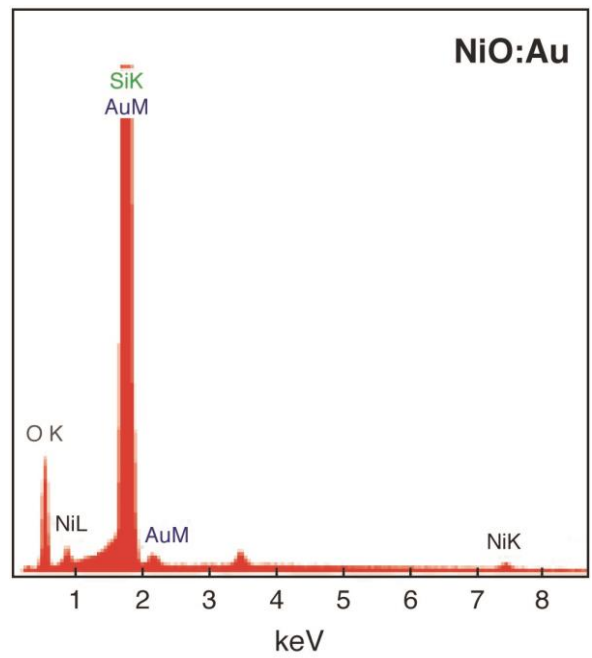

Figure 3

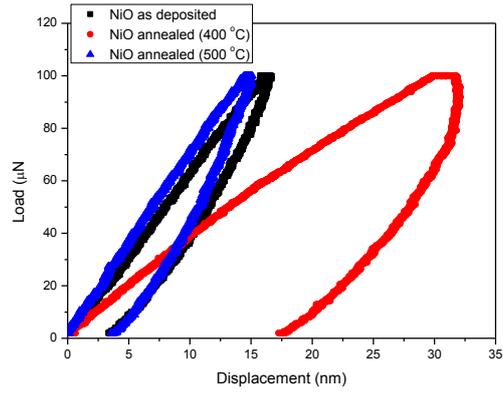

Figure 4

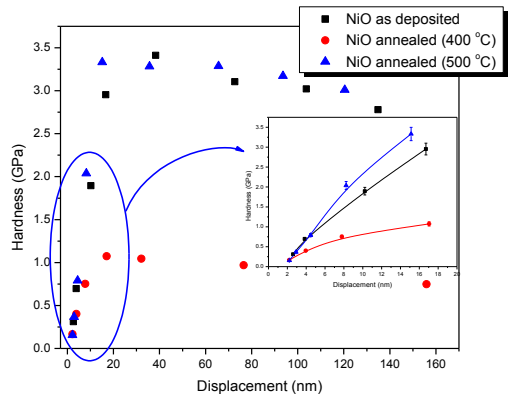

Figure 5a

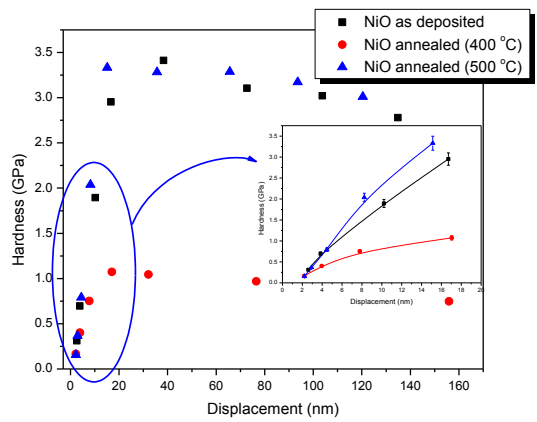

Figure 5b

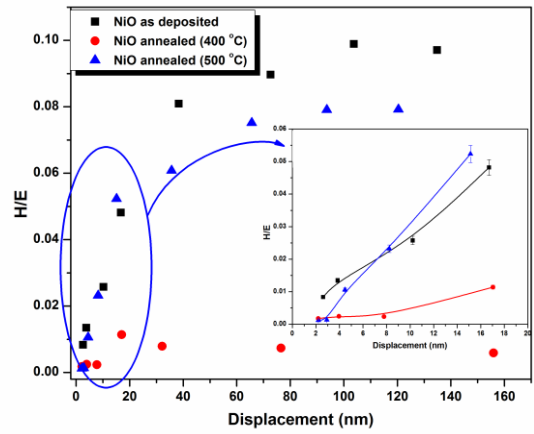

Figure 6a

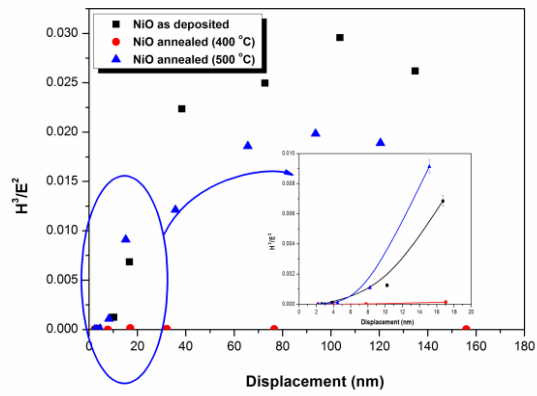

Figure 6b

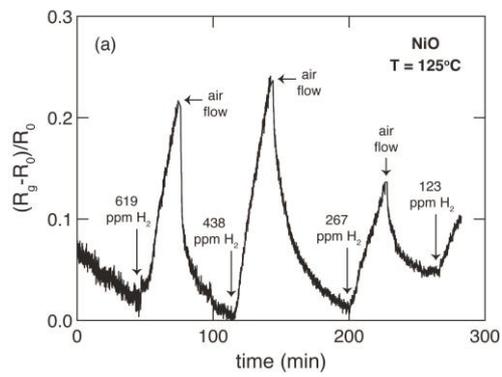

Figure 7a

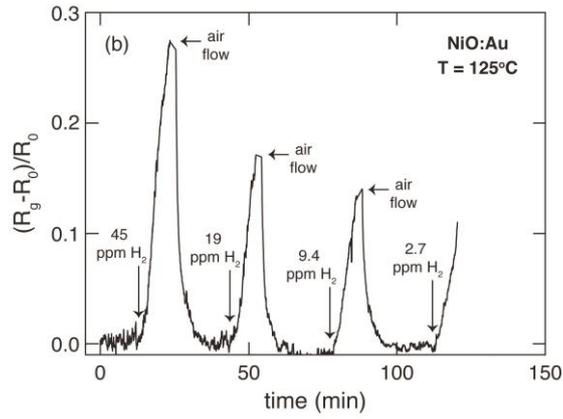

Figure 7b

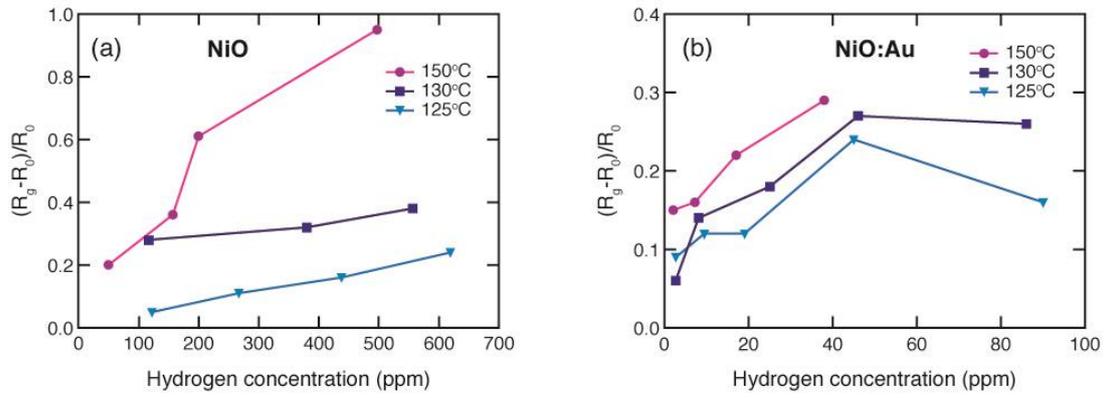

Figure 8

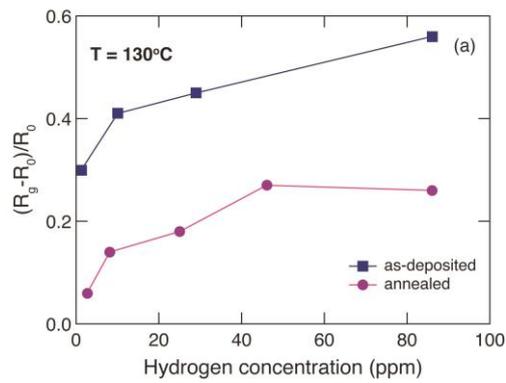

Figure 9a

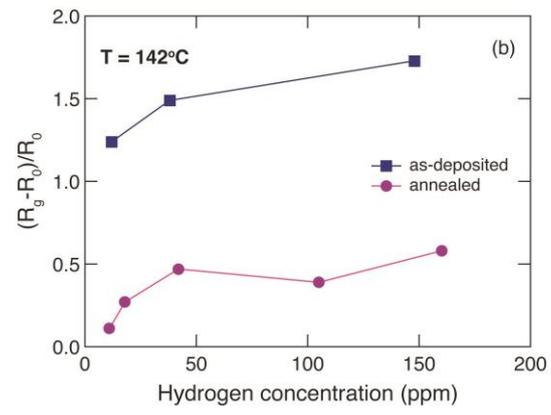

Figure 9b